\documentclass[preprint,aps,showpacs,preprintnumbers,amsmath,amsfonts]{revtex4}

\newcommand{\cC}{\ensuremath{\mathcal{C}}}
\newcommand{\cP}{\ensuremath{\mathcal{P}}}
\newcommand{\cT}{\ensuremath{\mathcal{T}}}
\newcommand{\cH}{\ensuremath{\mathcal{H}}}
\newcommand{\half}{\mbox{$\textstyle{\frac{1}{2}}$}}
\newcommand{\e}{{\rm e}}

\begin{document}

\title{The $\cC$ Operator in $\cP\cT$-Symmetric Quantum Field Theory Transforms
as a Lorentz Scalar}

\author{Carl M.\ Bender, Sebastian F.\ Brandt, Jun-Hua Chen, and Qinghai 
Wang}

\affiliation{Department of Physics, Washington University, St.\ Louis MO 63130,
USA}

\date{\today}

\begin{abstract}
A non-Hermitian Hamiltonian has a real positive spectrum and exhibits unitary
time evolution if the Hamiltonian possesses an unbroken $\cP\cT$ (space-time
reflection) symmetry. The proof of unitarity requires the construction of a
linear operator called $\cC$. It is shown here that $\cC$ is the complex
extension of the intrinsic parity operator and that the $\cC$ operator
transforms under the Lorentz group as a scalar.
\end{abstract}


\maketitle

\section{Introduction}
\label{s1}
If the Hamiltonian that defines a quantum theory is Hermitian $H=H^\dag$, then 
we can be sure that the energy spectrum of the Hamiltonian is real and that the
time evolution operator $U=\e^{iHt}$ is unitary (probability preserving). (The
symbol $\dag$ represents conventional Dirac Hermitian conjugation; that is,
combined complex conjugation and matrix transpose.) However, a non-Hermitian
Hamiltonian can also have these desired properties. For example, the quantum
mechanical Hamiltonians
\begin{equation}
H=p^2+x^2(ix)^\epsilon\quad(\epsilon\geq0)
\label{e1}
\end{equation}
have been shown to have real positive discrete spectra \cite{BB,DDT} and to
exhibit unitary time evolution \cite{BBJ}. Spectral positivity and unitarity
follow from a crucial symmetry property of the Hamiltonian; namely, $\cP\cT$
symmetry. (The linear operator $\cP$ represents parity reflection and the
anti-linear operator $\cT$ represents time reversal.) To summarize, if a
Hamiltonian has an unbroken $\cP\cT$ symmetry, then the energy levels are real
and the theory is unitary.

The proof of unitarity requires the construction of a new linear operator
called $\cC$, which in quantum mechanics is a sum over the eigenstates of the
Hamiltonian \cite{BBJ}. The $\cC$ operator is then used to define the inner
product for the Hilbert space of state vectors: $\langle A|B\rangle\equiv
A^{\cC\cP\cT}\cdot B$. With respect to this inner product, the time evolution
of the theory is unitary.

In the context of quantum mechanics, there have been a number of papers
published on the calculation of $\cC$. A perturbative calculation of $\cC$
in powers of $\epsilon$ for the cubic Hamiltonian
\begin{equation}
H=p^2+x^2+i\epsilon x^3
\label{e2}
\end{equation}
was performed \cite{BMW} and a perturbative calculation of $\cC$ for the
two- and three-degree-of-freedom Hamiltonians
\begin{equation}
H=p^2+q^2+x^2+y^2+i\epsilon xy^2
\label{e3}
\end{equation}
and
\begin{equation}
H=p^2+q^2+r^2+x^2+y^2+z^2+i\epsilon xyz
\label{e4}
\end{equation}
was also performed \cite{BX}. A nonperturbative WKB calculation of $\cC$ for the
quantum-mechanical Hamiltonians in (\ref{e1}) was done \cite{BJ}.

In quantum mechanics it was found that the $\cC$
operator has the general form
\begin{equation}
\cC=\e^Q \cP,
\label{e5}
\end{equation}
where $Q$ is a function of the dynamical variables (the coordinate and momentum
operators). The simplest way to calculate $\cC$ in non-Hermitian quantum
mechanics is to solve the three operator equations that define $\cC$:
\begin{equation}
\cC^2=\openone,\quad[\cC,\cP\cT]=0,\quad[\cC,H]=0.
\label{e6}
\end{equation}

Following successful calculations of the $\cC$ operator in quantum mechanics,
some calculations of the $\cC$ operator in field-theoretic models were done. A
leading-order perturbative calculation of $\cC$ for the self-interacting cubic
theory whose Hamiltonian density is
\begin{equation}
\cH({\bf x},t)=\half\pi^2({\bf x},t)+\half\mu^2\varphi^2({\bf x},t)+
\half[\nabla_{\!\bf x}\varphi({\bf x},t)]^2+i\epsilon\varphi^3({\bf x},t)
\label{e7}
\end{equation}
was performed \cite{Field,FField} and the $\cC$ operator was also calculated for
a $\cP\cT$-symmetric version of quantum electrodynamics \cite{FFField}. For each
of these field theory calculations, it was assumed that the form for the $\cC$
operator is that given in (\ref{e5}), where $\cP$ is now the field-theoretic
version of the parity reflection operator; that is, for a scalar field
\begin{equation}
\cP\varphi({\bf x},t)\cP = \varphi(-{\bf x},t),
\label{e8}
\end{equation}
and for a pseudoscalar field
\begin{equation}
\cP\varphi({\bf x},t)\cP=-\varphi(-{\bf x},t).
\label{e9}
\end{equation}

However, in a recent investigation of the Lee model in which we calculated the
$\cC$ operator exactly \cite{Lee}, we found that the correct field-theoretic
form for the $\cC$ operator is not $\cC=\e^Q\cP$ but rather
\begin{equation}
\cC=\e^Q \cP_I.
\label{e10}
\end{equation}
Here, $\cP_I$ is the {\it intrinsic} parity operator; $\cP_I$ has the same
effect on the fields as $\cP$ except that it does not change the sign of the
spatial argument of the fields. Thus, for a scalar field
\begin{equation}
\cP_I\varphi({\bf x},t)\cP_I = \varphi({\bf x},t),
\label{e11}
\end{equation}
and for a pseudoscalar field
\begin{equation}
\cP_I\varphi({\bf x},t)\cP_I=-\varphi({\bf x},t).
\label{e12}
\end{equation}

The fundamental difference between the conventional parity operator $\cP$ and
the intrinsic parity operator $\cP_I$ is seen in their Lorentz transformation
properties. For a quantum field theory that has parity symmetry the intrinsic
parity operator $\cP_I$ is a {\it Lorentz scalar} because $\cP_I$ commutes with
the generators of the homogeneous Lorentz group:
\begin{equation}
\left[\cP_I,J^{\mu\nu}\right]=0.
\label{e13}
\end{equation}
However, the conventional parity operator $\cP$ does not commute with a Lorentz
boost,
\begin{equation}
\left[\cP,J^{0i}\right]=-2J^{0i}\cP,
\label{e14}
\end{equation}
and $\cP$ transforms as an infinite-dimensional reducible representation of the
Lorentz group \cite{Parity}. Specifically, $\cP$ transforms as an infinite
direct sum of finite-dimensional tensors:
\begin{equation}
(0,1)\oplus(0,3)\oplus(0,5)\oplus(0,7)\oplus\cdots.
\label{e15}
\end{equation}
That is, $\cP$ transforms as a scalar plus the spin-$0$ component of a two-index
symmetric traceless tensor plus the spin-$0$ component of a four-index symmetric
traceless tensor plus the spin-$0$ component of a six-index symmetric traceless
tensor, and so on. Note that in (\ref{e15}) we use the notation of
Ref.~\cite{GMS}. It was shown in Ref.~\cite{Parity} that the decomposition in
(\ref{e15}) corresponds to a completeness summation over Wilson polynomials
\cite{Parity,KS}.

We believe that the correct way to represent the $\cC$ operator is in
(\ref{e10}) and not in (\ref{e5}).  In the case of quantum mechanics there is,
of course, no difference between these two representations because in this case
$\cP=\cP_I$. However, in quantum field theory, where $\cP\neq\cP_I$, these two
representations are different. For the case of the Lee model, (\ref{e5}) is the
wrong representation for $\cC$ and (\ref{e10}) is the correct representation. It
is most remarkable that for the case of the cubic quantum field theory in
(\ref{e7}) in either representation the functional $Q$ is exactly the same.
However, the representation of $\cC$ in (\ref{e10}) is strongly preferred
because, as we will show in this paper, it transforms as a Lorentz scalar.

The work in this paper indicates for the first time the physical and
mathematical interpretation of the $\cC$ operator: {\it The} $\cC$ {\it operator
is the complex analytic continuation of the intrinsic parity operator.} To
understand this remark, consider the Hamiltonian $H$ associated with the
Hamiltonian density $\cH$ in (\ref{e7}), $H=\int d{\bf x}\,\cH({\bf x},t)$.
When $\epsilon=0$, $H$ commutes with $\cP_I$ and $\cP_I$ transforms as a Lorentz
scalar. When $\epsilon\neq0$, $H$ does not commute with $\cP_I$, but 
$H$ {\it does} commute with $\cC$. Moreover, we will see that $\cC$
transforms as a Lorentz scalar. Furthermore, since $Q\to0$ as $\epsilon\to0$, we
see that $\cC\to\cP_I$ in this limit. Therefore, we can interpret the $\cC$
operator as the complex extension of the intrinsic parity operator.

This paper is organized very simply: In Sec.~\ref{s2} we examine an exactly
solvable $\cP\cT$-symmetric model quantum field theory and we calculate the
$\cC$ operator exactly. We show that this $\cC$ operator transforms as a
Lorentz scalar. Next, in Sec.~\ref{s3} we examine the cubic quantum theory whose
Hamiltonian density is given in (\ref{e7}) and show that the $\cC$ operator,
which was calculated to leading order in $\epsilon$ in Ref.~\cite{Field}, again
transforms as a Lorentz scalar. We make some concluding remarks in
Sec.~\ref{s4}.

\section{The $\cC$ Operator for a Toy Model Quantum Field Theory}
\label{s2}
In this section we calculate the $\cC$ operator for an exactly solvable
non-Hermitian $\cP\cT$-symmetric quantum field theory and show that $\cC$
transforms as a Lorentz scalar. Consider the theory defined by the
Hamiltonian density
\begin{equation}
\cH({\bf x},t)=\cH_0({\bf x},t)+ \epsilon \cH_1({\bf x},t),
\label{e16}
\end{equation}
where
\begin{equation}
\cH_0({\bf x},t)=\half\pi^2({\bf x},t)+\half\mu^2\varphi^2({\bf x},t)+\half[
\nabla_{\!\bf x}\varphi({\bf x},t)]^2,\qquad
\cH_1({\bf x},t)=i\varphi({\bf x},t).
\label{e17}
\end{equation}

We assume that the field $\varphi({\bf x},t)$ in (\ref{e17}) is a pseudoscalar.
Thus, under intrinsic parity reflection $\varphi({\bf x},t)$ transforms as in
(\ref{e12}). Also, $\pi({\bf x},t)={\dot \varphi}({\bf x},t)$, which is the
field that is dynamically conjugate to $\varphi({\bf x},t)$, transforms as
\begin{equation}
\cP_I\pi({\bf x},t)\cP_I=-\pi({\bf x},t).
\label{e18}
\end{equation}

To calculate the $\cC$ operator we assume that $\cC$ has the form in
(\ref{e10}). The operator $Q[\pi,\varphi]$, which is a functional of the fields
$\varphi$ and $\pi$, can be expressed as a series in powers of $\epsilon$:
\begin{equation}
Q=\epsilon Q_1+\epsilon^3Q_3+\epsilon^5Q_5+\cdots.
\label{e19}
\end{equation}
Note that only odd powers appear in the expansion of $Q$. As shown in
Ref.~\cite{Field}, we obtain $Q_1$ by solving the operator equation
\begin{equation}
\left[Q_1,\int\!\! d{\bf x}\, \cH_0({\bf x},t)\right] = 2 \int\!\! d{\bf 
x}\, \cH_1({\bf x},t).
\label{e20}
\end{equation}

To solve (\ref{e20}) we recall from Ref.~\cite{Field} that the functional $Q[\pi
,\varphi]$ is odd in $\pi$ and even in $\varphi$. We then substitute the
elementary {\it ansatz}
\begin{equation}
Q_1=\int\!\!d{\bf x}\,R({\bf x})\pi({\bf x},t),
\label{e21}
\end{equation}
where $R({\bf x})$ is a c-number. We find that $R({\bf x})$ satisfies the
functional equation
\begin{equation}
\int\!\!d{\bf x}\,\varphi({\bf x},t)(\mu^2-\nabla_{\!\bf x}^2)R({\bf 
x})=-2 \int\!\! d{\bf x}\, \varphi({\bf x},t),
\label{e22}
\end{equation}
whose solution is a constant:
\begin{equation}
R({\bf x})=-2/\mu^2.
\label{e23}
\end{equation}

All higher-order contributions to $Q[\pi,\varphi]$ in (\ref{e19}) vanish because
\begin{eqnarray}
\left[Q_1,\left[Q_1,H_1\right]\right]=\left[Q_1,-2V/\mu^2\right]=0,
\label{e24}
\end{eqnarray}
where $V$ is the volume of the space. Therefore, the sequence of equations in
Eq.~(34) of Ref.~\cite{Field} terminates after the first equation. We
thus obtain the exact result
\begin{equation}
\cC = \exp\left(-\frac{2\epsilon}{\mu^2}\int\!\!d{\bf x}\,\pi({\bf 
x},t)\right)\cP_I.
\label{e25}
\end{equation}

The $\cC$ operator is clearly a rotational scalar. To prove that $\cC$
transforms as a Lorentz scalar, we must show that $\cC$ commutes with the
Lorentz boost operator
\begin{equation}
J^{0i}=J^{0i}_0+\epsilon J^{0i}_1,
\label{e26}
\end{equation}
where
\begin{eqnarray}
J^{0i}_0(t) &=& t \int\!\! d{\bf x}\, \pi({\bf x},t) \nabla_{\!\bf x}^i 
\varphi({\bf x},t)-\int\!\! d{\bf x}\, x^i \cH_0({\bf x},t),\nonumber\\
J^{0i}_1(t) &=& -\int\!\! d{\bf x}\, x^i \cH_1({\bf x},t).
\label{e27}
\end{eqnarray}
The commutator of $\cC=\e^Q\cP_I$ with $J^{0i}$ in (\ref{e26}) has two terms:
\begin{eqnarray}
[\cC,J^{0i}]
&=& \left[\e^Q\cP_I,J^{0i}_0 + \epsilon J^{0i}_1\right]\nonumber\\
&=& \e^Q\left[\cP_I,J^{0i}_0 + \epsilon J^{0i}_1\right] + 
\left[\e^Q,J^{0i}_0 + \epsilon J^{0i}_1\right]\cP_I\nonumber\\
&=& -2\epsilon\,\e^QJ^{0i}_1\cP_I + \left[\e^Q,J^{0i}_0 + \epsilon 
J^{0i}_1\right]\cP_I.
\label{e28}
\end{eqnarray}
To evaluate the second term in (\ref{e28}) we use the general formula
\begin{equation}
\Big[f[\pi],\varphi({\bf x},t)\Big]=-i\frac{\delta}{\delta\pi({\bf x},t)}f[\pi],
\label{e29}
\end{equation}
from which we obtain
\begin{eqnarray}
\left[\exp\left(-\frac{2\epsilon}{\mu^2}\int\!\!d{\bf x}\,\pi({\bf x},t)\right),
\varphi({\bf x},t)\right]&=& -i\frac{\delta}{\delta\pi({\bf x},t)} \exp\left(-
\frac{2\epsilon}{\mu^2}\int\!\!d{\bf x}\,\pi({\bf x},t)\right)\nonumber\\
&=& \frac{2i\epsilon}{\mu^2}\exp\left(-\frac{2\epsilon}{\mu^2}\int\!\!d{\bf 
x}\,\pi({\bf x},t)\right).
\label{e30}
\end{eqnarray}
Thus, the second term in (\ref{e28}) evaluates to
\begin{eqnarray}
\left[\e^Q,J^{0i}_0 + \epsilon J^{0i}_1\right]
&=& t\int\!\! d{\bf x}\,\pi({\bf x},t)\nabla_{\!\bf x}^i \frac{2i\epsilon}{\mu^2
}\e^Q-\int\!\! d{\bf x}\, x^i\left(\frac{1}{2}\mu^2\varphi({\bf x},t)\frac{2i
\epsilon}{\mu^2}\e^Q+\frac{1}{2}\mu^2\frac{2i\epsilon}{\mu^2}\e^Q\varphi({\bf 
x},t)\right.\nonumber\\
&&\quad \left.+\frac{1}{2}\nabla_{\!\bf x} \varphi({\bf x},t)\cdot\nabla_{\!\bf
x}\frac{2i\epsilon}{\mu^2}\e^Q+\frac{1}{2}\nabla_{\!\bf x}\frac{2i\epsilon}
{\mu^2}\e^Q\cdot\nabla_{\!\bf x} \varphi({\bf x},t)+i\epsilon\frac{2i\epsilon}
{\mu^2}\e^Q\right)\nonumber\\
&=& -i\epsilon\int\!\!d{\bf x}\,x^i\varphi({\bf x},t)\e^Q-i\epsilon\e^Q\int\!\!
d{\bf x}\, x^i \varphi({\bf x},t)\nonumber\\
&=& -2i\epsilon\, \e^Q\int\!\! d{\bf x}\, x^i \varphi({\bf x},t).
\label{e31}
\end{eqnarray}
In the calculation above we assume that the integral $\int d{\bf x}\, x^i$
vanishes by oddness, and in the last step we use the identity
\begin{eqnarray}
\left[\int\!\!d{\bf x}\,x^i\varphi({\bf x},t),\int\!\!d{\bf y}\,\pi({\bf y},t)
\right]=i\int\!\!\!\!\int\!\!d{\bf x}\,d{\bf y}\,x^i\delta({\bf x}-{\bf y})
=i\int\!\! d{\bf x}\, x^i=0.
\label{e32}
\end{eqnarray}
Plugging (\ref{e31}) into the commutator (\ref{e28}), we obtain
\begin{eqnarray}
[\cC,J^{0i}]=0,
\label{e33}
\end{eqnarray}
which verifies that the $\cC$ operator for this toy model transforms as a
Lorentz scalar.

\section{The $\cC$ Operator for an $i\varphi^3$ Quantum Field Theory}
\label{s3}
We consider next the cubic quantum field theory described by the Hamiltonian
density in (\ref{e7}) and we rewrite $\cH$ as $\cH({\bf x},t)=\cH_0({\bf x},t)+ 
\epsilon \cH_1({\bf x},t)$, where
\begin{equation}
\cH_0({\bf x},t)=\half\pi^2({\bf x},t)+\half\mu^2\varphi^2({\bf x},
t)+\half[\nabla_{\!\bf x}\varphi({\bf x},t)]^2,\qquad
\cH_1({\bf x},t)=i\varphi^3({\bf x},t).
\label{e35}
\end{equation}
For this nontrivial field theory the $\cC$ operator still has the form $\cC=\e^Q
\cP_I$, where $Q$ is a series in odd powers of $\epsilon$, $Q=\epsilon Q_1+
\epsilon^3Q_3+\epsilon^5Q_5+\cdots$, and $\cP_I$ is the intrinsic parity
operator.

We expect that the operator $\cC$ is a rotational scalar because neither $Q$ nor
$\cP_I$ depend on spatial coordinates and one can verify that $\cC$ is indeed a
rotational scalar by showing that $\cC$ commutes with the generator of spatial
rotations $J^{ij}$. To prove that $\cC$ is a Lorentz scalar, we must show that
$\cC$ also commutes with the Lorentz boost operator $J^{0i}=J^{0i}_0+\epsilon
J^{0i}_1$, where the general formulas for $J^{0i}_0$ and $J^{0i}_1(t)$ are given
in (\ref{e27}).

Let us expand the commutator $[\cC,J^{0i}]$ in powers of $\epsilon$. To order
$\epsilon^2$, we have
\begin{eqnarray}
[\cC,J^{0i}]
&=& \left[\left(1+\epsilon Q_1+\half\epsilon^2 Q_1^2\right)\cP_I,J^{0i}_0+
\epsilon J^{0i}_1\right]+{\rm O}(\epsilon^3)\nonumber\\
&=& \left[\cP_I,J^{0i}_0\right] + \epsilon\left(\left[Q_1,J_0^{0i}\right]\cP_I+
Q_1\left[\cP_I,J_0^{0i}\right] +\left[\cP_I,J_1^{0i}\right]\right)\nonumber\\
&&\hspace{-.4cm}+\epsilon^2\left(\half \left[Q_1^2,J^{0i}_0\right]\cP_I+ \half 
Q_1^2\left[\cP_I,J^{0i}_0\right]+\left[Q_1,J^{0i}_1\right]\cP_I + 
Q_1\left[\cP_I,J^{0i}_1\right] \right)+{\rm O}(\epsilon^3).
\label{e36}
\end{eqnarray}
The leading term vanishes because $\cP_I$ commutes with $J^{0i}_0$: $\left[\cP_I
,J^{0i}_0\right]=0$. Using the identity $\left[\cP_I,J^{0i}_1\right]=-2J^{0i}_1
\cP_I$, we simplify (\ref{e36}) to
\begin{eqnarray}
[\cC,J^{0i}]
&=&\epsilon\left(\left[Q_1,J^{0i}_0\right]\cP_I-2J^{0i}_1\cP_I\right)\nonumber\\
&&\hspace{-.4cm}+\epsilon^2\left(\half Q_1\left[Q_1,J^{0i}_0\right]\cP_I+\half
\left[Q_1,J^{0i}_0\right]Q_1\cP_I-Q_1J^{0i}_1\cP_I-J^{0i}_1Q_1\cP_I\right)+
{\rm O}(\epsilon^3).
\label{e37}
\end{eqnarray}
Evidently, if the term of order $\epsilon$ vanishes, then the $\epsilon^2$ 
term vanishes automatically.

From (\ref{e37}) we can see that to prove that $\cC$ is a scalar we need to show
that $\left[Q_1,J^{0i}_0\right]=2J^{0i}_1$. The generator $J_0^{0i}$ in
(\ref{e27}) consists of two parts, and the first part commutes with  $Q_1$. To
show that this is so, we argue that for any functional of $\pi$ and $\varphi$,
say $f[\pi,\varphi]$, the first term of $J^{0i}_0$ in (\ref{e27}) commutes with
$f[\pi,\varphi]$: $\left[f[\pi,\varphi],t\int d{\bf x}\,\pi({\bf x},t)\nabla_{\!
\bf x}^i\varphi({\bf x},t)\right]=0$. We can verify this either by using
time-translation invariance and setting $t=0$ or by noting that this commutator
is explicitly an integral of a total derivative:
\begin{eqnarray}
&&\left[f[\pi,\varphi],t \int d{\bf y}\, \pi({\bf y},t) \nabla_{\!\bf 
y}^i \varphi({\bf y},t)\right]\nonumber\\
&=& t \int d{\bf y}\, \left(i \frac{\delta}{\delta \varphi({\bf 
y},t)}f[\pi,\varphi]\nabla_{\!\bf y}^i \varphi({\bf y},t)- i \pi({\bf y},t)
\nabla_{\!\bf y}^i \frac{\delta}{\delta \pi({\bf y},t)}f[\pi,\varphi]\right),
\label{e38}
\end{eqnarray}
where we have used the variational formulas $\big[f\left[\pi,\varphi\right],
\varphi({\bf x},t)\big]=-i\frac{\delta}{\delta\pi({\bf x},t)}f[\pi,\varphi]$ and
$\big[f\left[\pi,\varphi\right],\pi({\bf x},t)\big]=i\frac{\delta}{\delta 
\varphi({\bf x},t)}f[\pi,\varphi]$. Integrating the second term of (\ref{e38})
by parts, we get
\begin{eqnarray}
&&\left[f[\pi,\varphi],t \int d{\bf y}\, \pi({\bf y},t) \nabla_{\!\bf 
y}^i \varphi({\bf y},t)\right]\nonumber\\
&=& t \int d{\bf y}\left(i \frac{\delta}{\delta \varphi({\bf 
y},t)}f[\pi,\varphi] \nabla_{\!\bf y}^i \varphi({\bf y},t)+ i 
\nabla_{\!\bf y}^i \pi({\bf y},t) \frac{\delta}{\delta \pi({\bf 
y},t)}f[\pi,\varphi]\right)=0.
\label{e39}
\end{eqnarray}
Thus, since $Q_1$ is a functional of $\pi$ and $\varphi$, only the second term
of $J_0^{0i}$ in (\ref{e27}) contributes to the commutator of $Q_1$ with $J^{0i}
_0$: $\left[Q_1,J^{0i}_0\right]=-\left[Q_1,\int d{\bf x}\,x^i\cH_0({\bf x},t)
\right]$.

Thus, we have reduced the problem of showing that $\cC$ is a scalar to
establishing the commutator identity
\begin{equation}
\left[Q_1,\int d{\bf x}\,x^i\Big(\half\pi^2({\bf x},t)+\half\mu^2\varphi^2({\bf
x},t)+\half[\nabla_{\!\bf x}\varphi({\bf x},t)]^2\Big)\right]=2i\int d{\bf x}\,
x^i \varphi^3({\bf x},t).
\label{e40}
\end{equation}
This equation is similar in structure to Eq.~(66) in Ref.~\cite{Field},
\begin{eqnarray}
\left[Q_1,\int\!\!d{\bf x}\,\Big(\half\pi^2({\bf x},t)+\half\mu^2\varphi^2({\bf
x}, t)+\half[\nabla_{\!\bf x}\varphi({\bf x},t)]^2\Big)\right]=
2i\int\!\!d{\bf x}\,\varphi^3({\bf x},t),
\label{e41}
\end{eqnarray}
apart from an integration by parts. The only difference between (\ref{e40}) and
(\ref{e41}) is that there are extra factors of $x^i$ in the integrands of
(\ref{e40}).

We will now follow the same line of analysis used in Ref.~\cite{Field} to solve
(\ref{e40}). We introduce the identical {\it ansatz} for $Q_1$:
\begin{eqnarray}
Q_1=\int\!\!\!\!\int\!\!\!\!\int\!\!d{\bf x}\,d{\bf y}\,d{\bf z}\,M_{({\bf xyz})
}\pi_{\bf x}\pi_{\bf y}\pi_{\bf z}+\int\!\!\!\!\int\!\!\!\!\int\!\!d{\bf x}\,
d{\bf y}\,d{\bf z}\,N_{{\bf x}({\bf yz})}\varphi_{\bf y}\pi_{\bf x}\varphi_{\bf
z},
\label{e42}
\end{eqnarray}
where we have suppressed the time variable $t$ in the fields and have indicated
spatial dependences with subscripts. To indicate that the unknown function $M$
is totally symmetric in its three arguments, we use the notation $M_{({\bf x}
{\bf y}{\bf z})}$ and we write $N_{{\bf x}({\bf yz})}$ because the unknown
function $N$ is symmetric under the interchange of the second and third
arguments.

Performing the commutator in (\ref{e40}), we obtain two functional equations:
\begin{eqnarray}
\int\!\!\!\!\int\!\!\!\!\int\!\!d{\bf x}\,d{\bf y}\,d{\bf 
z}\,\varphi_{\bf x}\varphi_{\bf y}\varphi_{\bf 
z}\left[(x^i+y^i+z^i)\mu^2-\nabla_{\!\bf x}^i - x^i\nabla_{\!\bf 
x}^2\right]N_{{\bf x}({\bf y}{\bf z})}=-2\int\!\!d{\bf w}\,w^i\varphi_{\bf w}^3,
\label{e43}
\end{eqnarray}
\begin{eqnarray}
&&\hspace{-0.6cm}\int\!\!\!\!\int\!\!\!\!\int\!\!d{\bf x}\,d{\bf y}\,d{\bf z}\,
\left(y^i\pi_{\bf y}\pi_{\bf x}\varphi_{\bf z}+z^i\varphi_{\bf y}\pi_{\bf x}
\pi_{\bf z}\right)N_{{\bf x}({\bf yz})}\nonumber\\
&&~~~=\int\!\!\!\!\int\!\!\!\!\int\!\!d{\bf x}\,d{\bf y}\, d{\bf z}\,\Bigl[
\varphi_{\bf x}\pi_{\bf y}\pi_{\bf z}\left(x^i\mu^2-\nabla_{\!\bf x}^i-x^i\nabla
_{\!\bf x}^2\right)M_{({\bf xyz})}\nonumber\\
&& \quad +\pi_{\bf x}\varphi_{\bf y}\pi_{\bf z}\left(y^i\mu^2-\nabla_{\!\bf y}^i
- y^i\nabla_{\!\bf y}^2\right)M_{({\bf xyz})}+\pi_{\bf x}\pi_{\bf y}\varphi_{\bf
z}\left(z^i\mu^2-\nabla_{\!\bf z}^i-z^i\nabla_{\!\bf z}^2\right)M_{({\bf xyz})}
\Bigr].
\label{e44}
\end{eqnarray}
Next, we commute (\ref{e43}) three times with $\pi$ and commute (\ref{e44}) once
with $\pi$ and twice with $\varphi$ to transform these operator identities into
coupled differential equations for $M$ and $N$:
\begin{eqnarray}
&&\left[x^i(\mu^2-\nabla_{\!\bf x}^2)-\nabla_{\!\bf x}^i\right]N_{{\bf 
x}({\bf yz})}
+\left[y^i(\mu^2-\nabla_{\!\bf y}^2)-\nabla_{\!\bf y}^i\right] N_{{\bf 
y}({\bf xz})}
+\left[z^i(\mu^2-\nabla_{\!\bf z}^2)-\nabla_{\!\bf z}^i\right] N_{{\bf 
z}({\bf xy})}\nonumber\\
&&\qquad\qquad=-6x^i\delta({\bf x}-{\bf y})\delta({\bf x}-{\bf z}),
\label{e45}
\end{eqnarray}
\begin{eqnarray}
z^iN_{{\bf y}({\bf x}{\bf z})}+y^iN_{{\bf z}({\bf x}{\bf y})}=3\left[x^i(\mu^2-
\nabla_{\!\bf x}^2)-\nabla_{\!\bf x}^i\right]M_{({\bf xyz})}.
\label{e46}
\end{eqnarray}

Equation (\ref{e45}) is similar to Eq.~(71) in Ref.~\cite{Field}:
\begin{equation}
(\mu^2-\nabla_{\!\bf x}^2)N_{{\bf x}({\bf yz})}+(\mu^2-\nabla_{\!\bf y}^2)
N_{{\bf y}({\bf xz})}+(\mu^2-\nabla_{\!\bf z}^2)N_{{\bf z}({\bf xy})}
=-6\delta({\bf x}-{\bf y})\delta({\bf x}-{\bf z}).
\label{e71old}
\end{equation}
By permuting ${\bf x}$, ${\bf y}$ and ${\bf z}$, we rewrite Eq.~(72) of
Ref.~\cite{Field} as
\begin{equation}
N_{{\bf y}({\bf x}{\bf z})}+N_{{\bf z}({\bf x}{\bf y})}=3(\mu^2-\nabla_{\!\bf x}
^2)M_{({\bf xyz})}.
\label{e55}
\end{equation}
This equation is similar to Eq.~(\ref{e46}). The solutions for $M$ and $N$ are given in Eqs.~(83) and (84) of Ref.~\cite{Field}:
\begin{eqnarray}
M_{({\bf xyz})}=-\frac{4}{(2\pi)^{6}}\!\int\!\!\!\!\int\!\!d{\bf p}\,d{\bf q}
\,\frac{e^{i({\bf x}-{\bf y})\cdot{\bf p}+i({\bf x}-{\bf z})\cdot{\bf q}}}
{{\mathcal{D}}({\bf p},{\bf q})},
\label{e83old}
\end{eqnarray}
where ${\mathcal{D}}({\bf p},{\bf q})=4[{\bf p}^2{\bf q}^2-({\bf p}\cdot{\bf q})
^2]+4\mu^2({\bf p}^2+{\bf p}\cdot{\bf q}+{\bf q}^2)+3\mu^4$, and
\begin{eqnarray}
N_{{\bf x}({\bf yz})} &=& 3\left(\nabla_{\!\bf y}\cdot\nabla_{\!\bf z}+\half\mu^2
\right)M_{({\bf xyz})}\nonumber\\
&=& -\frac{12}{(2\pi)^6}\!\int\!\!\!\!\int\!\!d{\bf p}\,d{\bf
q}\,\frac{\left(-{\bf p}\cdot{\bf q}+\half\mu^2\right)\e^{i({\bf x}-{\bf y})
\cdot{\bf p}+i({\bf x}-{\bf z})\cdot{\bf q}}} {{\mathcal{D}}({\bf p},{\bf q})}.
\label{e48}
\end{eqnarray}
If these expressions for $M$ and $N$ solve (\ref{e45}) and (\ref{e46}), then we can claim that the operator
$\cC$ transforms as a scalar. To show that this is indeed true, we multiply
(\ref{e71old}) by $x^i$ and then subtract the result from (\ref{e45}) to obtain
\begin{equation}
(y^i-x^i)(\mu^2-\nabla_{\!\bf y}^2)N_{{\bf y}({\bf x}{\bf z})}+(z^i-x^i)(\mu^2-
\nabla_{\!\bf z}^2)N_{{\bf z}({\bf x}{\bf y})}=\nabla_{\!\bf x}^iN_{{\bf x}({\bf
y}{\bf z})}+\nabla_{\!\bf y}^iN_{{\bf y}({\bf x}{\bf z})}+\nabla_{\!\bf z}^iN_{
{\bf z}({\bf x}{\bf y})}.
\label{e47}
\end{equation}

By making the change of variables ${\bf p}\to{\bf q}$, ${\bf q}\to-{\bf p}-{\bf
q}$, we obtain $N_{{\bf y}({\bf xz})}$ in the form
\begin{eqnarray}
N_{{\bf y}({\bf xz})}=-\frac{12}{(2\pi)^6}\!\int\!\!\!\!\int\!\!d{\bf p}\,d{\bf
q}\,\frac{\left[{\bf q}\cdot({\bf p}+{\bf q})+\half\mu^2\right]\e^{i({\bf x}-{
\bf y})\cdot{\bf p}+i({\bf x}-{\bf z})\cdot{\bf q}}}{{\mathcal{D}}({\bf p},
{\bf q})},
\label{e49}
\end{eqnarray}
where we have used ${\mathcal{D}}({\bf q},-{\bf p}-{\bf q})={\mathcal{D}}({\bf
p},{\bf q})$. Similarly, by making the change of variables ${\bf p}\to-{\bf p}-
{\bf q}$, ${\bf q}\to{\bf p}$ and using the identity ${\mathcal{D}}(-{\bf p}-
{\bf q},{\bf p})={\mathcal{D}} ({\bf p},{\bf q})$, we obtain
\begin{eqnarray}
N_{{\bf z}({\bf xy})}=-\frac{12}{(2\pi)^6}\!\int\!\!\!\!\int\!\!d{\bf p}\,d{\bf
q}\,\frac{\left[{\bf p}\cdot({\bf p}+{\bf q})+\half\mu^2\right]\e^{i({\bf x}-{
\bf y})\cdot{\bf p}+i({\bf x}-{\bf z})\cdot{\bf q}}}{{\mathcal{D}}({\bf p},
{\bf q})}.
\label{e50}
\end{eqnarray}
Thus, from (\ref{e48}), (\ref{e49}), and (\ref{e50}) we find that the right-hand
side of (\ref{e47}) becomes
\begin{eqnarray}
{\rm RHS}~{\rm of}~(50)=\frac{12i}{(2\pi)^6}\!\int\!\!\!\!\int\!\!d{\bf p}\,d{
\bf q} \,\frac{\left[{\bf p}^i({\bf q}^2+2{\bf p}\cdot{\bf q})+{\bf q}^i({\bf 
p}^2+2{\bf p}\cdot{\bf q})\right]\e^{i({\bf x}-{\bf y})\cdot{\bf 
p}+i({\bf x}-{\bf z})\cdot{\bf q}}} {{\mathcal{D}}({\bf p},{\bf q})}.
\label{e51}
\end{eqnarray}
Also, substituting (\ref{e49}) and (\ref{e50}) into the LHS of (\ref{e47}), we
get
\begin{eqnarray}
{\rm LHS}~{\rm of}~(50)&=&-\frac{12}{(2\pi)^6}\!\int\!\!\!\!\int\!\!d{\bf p}\,d{
\bf q}\,\left\{(y^i-x^i)\left[{\bf q}\cdot({\bf p}+{\bf q})+\half\mu^2\right]({
\bf p}^2+\mu^2)\right.\nonumber\\
&&\qquad\left.+(z^i-x^i)\left[{\bf p}\cdot({\bf p}+{\bf q})+\half\mu^2\right]({
\bf q}^2+\mu^2)\right\}\frac{\e^{i({\bf x}-{\bf y})\cdot{\bf p}+i({\bf x}-{\bf z
})\cdot{\bf q}}}{{\mathcal{D}}({\bf p},{\bf q})}.
\label{e52}
\end{eqnarray}

We must now show that the RHS of (\ref{e47}) in (\ref{e51}) and the LHS of
(\ref{e47}) in (\ref{e52}) are equal. To do so, we substitute the identities
$(y^i-x^i)\e^{i({\bf x}-{\bf y})\cdot{\bf p}+i({\bf x}-{\bf z})\cdot{\bf q}}
=i\nabla_{\!\bf p}^i \e^{i({\bf x}-{\bf y})\cdot{\bf p}+i({\bf x}-{\bf z})\cdot
{\bf q}}$ and $(z^i-x^i)\e^{i({\bf x}-{\bf y})\cdot{\bf p}+i({\bf x}-{\bf z})
\cdot{\bf q}}=i\nabla_{\!\bf q}^i \e^{i({\bf x}-{\bf y})\cdot{\bf p}+i({\bf x}-
{\bf z})\cdot{\bf q}}$ into (\ref{e52}) and obtain
\begin{eqnarray}
{\rm LHS}~{\rm of}~(50)&=&-\frac{12i}{(2\pi)^6}\!\int\!\!\!\!\int\!\!d{\bf p}\,
d{\bf q}\,\left\{\frac{\left[{\bf q}\cdot({\bf p}+{\bf q})+\half\mu^2\right]({
\bf p}^2+\mu^2)}{{\mathcal{D}}({\bf p},{\bf q})}\nabla_{\!\bf p}^i\right.
\nonumber\\
&&\qquad\left.+\frac{\left[{\bf p}\cdot({\bf p}+{\bf 
q})+\half\mu^2\right]({\bf q}^2+\mu^2)}{{\mathcal{D}}({\bf p},{\bf 
q})}\nabla_{\!\bf q}^i\right\}\e^{i({\bf x}-{\bf y})\cdot{\bf p}+i({\bf 
x}-{\bf z})\cdot{\bf q}}.
\label{e53}
\end{eqnarray}
Integration by parts then yields
\begin{eqnarray}
{\rm LHS}~{\rm of}~(50)&=&\frac{12i}{(2\pi)^6}\!\int\!\!\!\!\int\!\!d{\bf p}\,
d{\bf q}\,\left\{\nabla_{\!\bf p}^i\frac{\left[{\bf q}\cdot({\bf p}+{\bf q})+
\half\mu^2\right]({\bf p}^2+\mu^2)}{{\mathcal{D}}({\bf p},{\bf q})}\right.
\nonumber\\
&&\qquad\left.+\nabla_{\!\bf q}^i\frac{\left[{\bf p}\cdot({\bf p}+{\bf q})+\half
\mu^2\right]({\bf q}^2+\mu^2)}{{\mathcal{D}}({\bf p},{\bf q})}\right\}\e^{i({\bf
x}-{\bf y})\cdot{\bf p}+i({\bf x}-{\bf z})\cdot{\bf q}}.
\label{e54}
\end{eqnarray}
Finally, we substitute the algebraic identity
\begin{eqnarray}
&&\nabla_{\!\bf p}^i\frac{\left[{\bf q}\cdot({\bf p}+{\bf 
q})+\half\mu^2\right]({\bf p}^2+\mu^2)}{{\mathcal{D}}({\bf p},{\bf q})}
+\nabla_{\!\bf q}^i\frac{\left[{\bf p}\cdot({\bf p}+{\bf 
q})+\half\mu^2\right]({\bf q}^2+\mu^2)}{{\mathcal{D}}({\bf p},{\bf 
q})}\nonumber\\
&=&\frac{\left[{\bf p}^i({\bf q}^2+2{\bf p}\cdot{\bf q})+{\bf q}^i({\bf 
p}^2+2{\bf p}\cdot{\bf q})\right]}{{\mathcal{D}}({\bf p},{\bf q})}\nonumber
\end{eqnarray}
to establish that the LHS and the RHS of (\ref{e47}) are equal. By combining
this result with (\ref{e71old}), we show that (\ref{e45}) holds.

Using the same technique, we can prove that the Eq.~(\ref{e46}) holds as well.
First, we multiply (\ref{e55}) by $x^i$ and subtract the result from (\ref{e46}) to get
\begin{equation}
(z^i-x^i)N_{{\bf y}({\bf x}{\bf z})}+(y^i-x^i)N_{{\bf z}({\bf x}{\bf y})}=-3
\nabla_{\!\bf x}^iM_{({\bf xyz})}.
\label{e56}
\end{equation}
Second, following the technique we used to derive (\ref{e54}), we find that the LHS of
(\ref{e56}) becomes
\begin{eqnarray}
{\rm LHS}~{\rm of}~(57)&=&\frac{12i}{(2\pi)^6}\!\int\!\!\!\!\int\!\!d{\bf p}\,
d{\bf q}\,\left[\nabla_{\!\bf q}^i\frac{{\bf q}\cdot({\bf p}+{\bf q})+\half\mu^2
}{{\mathcal{D}}({\bf p},{\bf q})}\right.\nonumber\\
&&\left.+\nabla_{\!\bf p}^i\frac{{\bf p}\cdot({\bf p}+{\bf q})+\half\mu^2}{{
\mathcal{D}}({\bf p},{\bf q})}\right]\e^{i({\bf x}-{\bf y})\cdot{\bf p}+i({\bf
x}-{\bf z})\cdot{\bf q}},
\label{e57}
\end{eqnarray}
and by substituting (\ref{e83old}) into (\ref{e56}), we find that
the RHS of (\ref{e56}) becomes
\begin{eqnarray}
{\rm RHS}~{\rm of}~(57) &=& \frac{12i}{(2\pi)^6}\!\int\!\!\!\!\int\!\!d{\bf p}
\,d{\bf q}\,\frac{({\bf p}^i+{\bf q}^i)\e^{i({\bf x}-{\bf y})\cdot{\bf p}+i({\bf
x}-{\bf z})\cdot{\bf q}}} {{\mathcal{D}}({\bf p},{\bf q})}.
\label{e58}
\end{eqnarray}
Finally, we substitute the algebraic identity
\begin{eqnarray}
\nabla_{\!\bf q}^i\frac{{\bf q}\cdot({\bf p}+{\bf q})+\half\mu^2}{{\mathcal{D}}
({\bf p},{\bf q})}+\nabla_{\!\bf p}^i\frac{{\bf p}\cdot({\bf p}+{\bf q})+\half
\mu^2}{{\mathcal{D}}({\bf p},{\bf q})}=\frac{{\bf p}^i+{\bf q}^i}{{\mathcal{D}}
({\bf p},{\bf q})}.
\nonumber
\end{eqnarray}
to establish that the LHS of (\ref{e56}) and the RHS of (\ref{e56}) are equal.
Thus, (\ref{e56}) holds and we may combine this result with (\ref{e55}) to establish that (\ref{e46}) is valid. We conclude that to order ${\rm O}(\epsilon^2)$ the operator $\cC$ transforms as a Lorentz scalar.

\section{Final Remarks}
\label{s4}
We have shown that the operator $\cC$ in a toy model $i\varphi$ quantum field
theory transforms as a Lorentz scalar and that to order ${\rm O}(\epsilon^2)$
the $\cC$ operator in an $i\epsilon\varphi^3$ quantum field theory transforms as
a Lorentz scalar. Based on this work, we conjecture that in general the $\cC$
operator for any unbroken $\cP\cT$-symmetric quantum field theory always
transforms as a Lorentz scalar.

By establishing the Lorentz transformation properties of $\cC$, we can now begin
to understand the role played by this mysterious operator: Apparently, the $\cC$
operator is the non-Hermitian $\cP\cT$-symmetric analog of the intrinsic parity
operator $\cP_I$. A conventional Hermitian quantum field theory does not have a
$\cC$ operator. However, if we begin with the Hermitian theory corresponding to
$\epsilon=0$ and turn on $\epsilon$, then the (scalar) $\cP_I$ symmetry of
Hermitian theory disappears and is replaced by the (scalar) $\cC$ symmetry of
the non-Hermitian theory.

The complex cubic quantum field theory discussed here is especially important
because it controls the dynamics of Reggeon field theory \cite{R} and describes
the Lee-Yang edge singularity \cite{LY}. The work in this paper shows that an $i
\varphi^3$ field theory is a consistent unitary quantum field theory on a
Hilbert space having a {\it Lorentz invariant} inner product.

\section*{Acknowledgements}
This work was supported in part by the U.S.~Department of Energy.


\begin{thebibliography}{99}

\bibitem{BB} C.~M.~Bender and S.~Boettcher, Phys.~Rev.~Lett.
{\bf 80}, 5243 (1998).

\bibitem{DDT} P.~Dorey, C.~Dunning and R.~Tateo, J.~Phys.~A {\bf 34} L391
(2001); {\em ibid}. {\bf 34}, 5679 (2001).

\bibitem{BBJ} C.~M.~Bender, D.~C.~Brody, and H.~F.~Jones,
Phys.~Rev.~Lett. {\bf 89}, 270402 (2002).

\bibitem{BMW} C.~M.~Bender, P.~N.~Meisinger, and Q.~Wang,
J.~Phys.~A {\bf 36}, 1973 (2003).

\bibitem{BX} C.~M.~Bender, J.~Brod, A.~Refig, and M.~E.~Reuter,
J.~Phys.~A: Math.~Gen.~{\bf 37}, 10139 (2004).

\bibitem{BJ} C.~M.~Bender and H.~F.~Jones, Phys.~Lett.~A {\bf 328}, 102 (2004).

\bibitem{Field} C.~M.~Bender, D.~C.~Brody, and H.~F.~Jones,
Phys.~Rev.~D {\bf 70}, 025001 (2004).

\bibitem{FField} C.~M.~Bender, D.~C.~Brody, and H.~F.~Jones,
Phys.~Rev.~Lett.~{\bf 93}, 251601 (2004).

\bibitem{FFField} C.~M.~Bender, I.~Cavero-Pelaez, K.~A.~Milton, and
K.~V.~Shajesh, manuscript in preparation.

\bibitem{Lee} C.~M.~Bender, S.~F.~Brandt, J.-H.~Chen, and Q.~Wang,
arXiv: hep-th/0411064.

\bibitem{Parity} C.~M.~Bender, P.~N.~Meisinger, and Q.~Wang, arXiv:
math-ph/0412001.

\bibitem{GMS} I.~M.~Gel'fand, R.~A.~Minlos, and Z.~Ya.~Shapiro, {\it
Representations of the Rotation and Lorentz Groups and Their
Applications} (MacMillan, New York, 1963).

\bibitem{KS} R.~Koekoek and R.~F.~Swarttouw, {\it The Askey-Scheme of
Hypergeometric Orthogonal Polynomials and its $q$-Analogue}
(http://aw.twi.tudelft.nl/$\sim$koekoek/askey/, 1998).

\bibitem{R} H.~D.~I.~Abarbanel, J.~D.~Bronzan, R.~L.~Sugar, and A.~R.~White,
Phys.~Rep.~{\bf 21}, 119 (1975); R.~Brower, M.~Furman, and M.~Moshe,
Phys.~Lett.~B {\bf 76}, 213 (1978).

\bibitem{LY} M.~E.~Fisher, Phys. Rev. Lett.~{\bf 40}, 1610 (1978); J.~L.~Cardy,
{\it ibid}.~{\bf 54}, 1345 (1985); J.~L.~Cardy and G. Mussardo, Phys.~Lett.~B
{\bf 225}, 275 (1989); A. B. Zamolodchikov, Nucl. Phys. B {\bf 348}, 619 (1991).

\end{thebibliography}
\end{document}